# In-situ 3D Imaging of Catalysis Induced Strain in Gold Nanoparticles

*Andrew Ulvestad[1,2]\*, Kiran Sasikumar[3], Jong Woo Kim[1,4], Ross Harder[4], Evan Maxey[4], Jesse N. Clark[5,6], Badri Narayanan[3], Sanket A. Deshmukh[3], Nicola Ferrier[7], Paul Mulvaney[8], Subramanian K.R.S. Sankaranarayanan[3], and Oleg G. Shpyrko[1].*

[1]Department of Physics, University of California-San Diego, La Jolla, California 92093-0319, USA

[2]Materials Science Division, Argonne National Laboratory, Argonne, Illinois 60439, USA

[3]Center for Nanoscale Materials, Argonne National Laboratory, Argonne, Illinois 60439, USA

[4]Advanced Photon Source, Argonne National Laboratory, Argonne, Illinois 60439, USA

[5]Stanford PULSE Institute, SLAC National Accelerator Laboratory Menlo Park, California 94025, USA

[6]Center for Free-Electron Laser Science (CFEL), Deutsches Elektronensynchrotron (DESY), Notkestrasse 85, 22607 Hamburg, Germany

[7]Mathematics and Computer Science Division, Argonne National Laboratory, Argonne, Illinois 60439, USA

[8]School of Chemistry & Bio21 Institute, University of Melbourne, Parkville, VIC 3010, Australia

**Corresponding Author**

*aulvestad@anl.gov

**Abstract**

Multi-electron transfer processes, such as hydrogen and oxygen evolution reactions, are crucially important in energy and biological science but require favorable catalysts to achieve fast kinetics. Nanostructuring catalysts can dramatically improve their properties, which can be difficult to understand due to strain and size dependent thermodynamics, the influence of defects, and substrate dependent activities. Here, we report 3D imaging of single gold nanoparticles during catalysis of ascorbic acid decomposition using Bragg coherent diffractive imaging (BCDI) as a route to eliminate ensemble effects while elucidating the strain-activity connection. Local strains were measured in single nanoparticles and modeled using reactive molecular dynamics (RMD) simulations and finite element analysis (FEA) simulations. RMD reveals a new chemical pathway for local strain generation in the gold lattice: chemisorption of hydroxyl ions. FEA reveals that the RMD results are transferable to the larger nanocrystal sizes studied in the experiment. Our study reveals the strain-activity connection and opens a powerful new avenue for joint theoretical and experimental studies of multi-electron transfer processes catalyzed by nanocrystals.

**TOC GRAPHICS**:

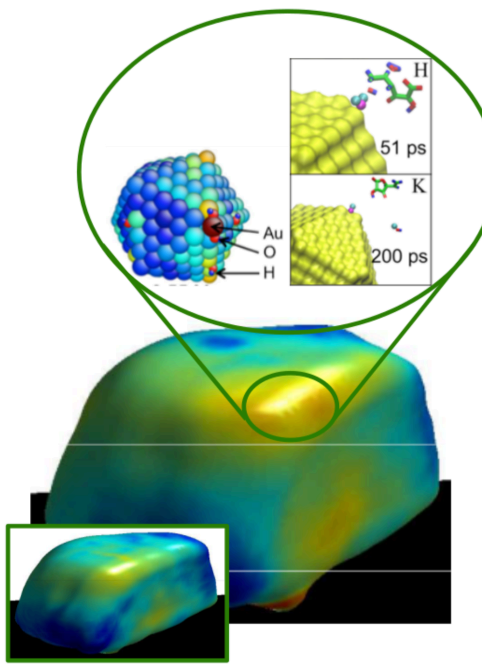

Investigating the catalytic activity of nanostructures relevant to multi-electron transfer processes is an important area of research with both energy and biological implications[1–3]. Though bulk gold is chemically inert, it was recently shown that gold nanoparticles are catalytically active in a diverse set of reactions[4,5], including ascorbic acid (Vitamin C) decomposition, an important biological process. Gold nanoparticles have been studied by a variety of techniques, including spectroscopies such as surface plasmon[6,7] and Fourier transform infrared[8], in addition to electron microscopy[9,10]. However, these techniques can be challenging to use in reactive environments and are not directly sensitive to the 3D strain field inside the nanocrystal, which can significantly affect catalytic activity[11,12]. Recently, experimental techniques have evolved to conduct time-dependent lattice dynamics measurements in nanomaterials[13–17] via Bragg coherent diffractive imaging (BCDI) measurements with sub-twenty nanometer resolution. Integrating 3D displacement and strain field measurements with suitable simulation techniques can provide crucial physical insights, including linking the strain field with chemistry[18,19]. An understanding of *in-situ* 3D lattice dynamics could aid in developing improved catalysts in the future.

In this work, we study ascorbic acid decomposition catalyzed by a gold nanoparticle as a model multi-electron transfer process to investigate the connection between strain and catalysis. BCDI measurements reveal reversible lattice distortions in gold nanocrystals upon exposure to ascorbic acid solutions. We employ RMD to uncover the reaction pathway and the origin of lattice strain. Finally, RMD-aided FEA is used to connect the experimental and theoretical results.

Bragg coherent diffractive imaging is a powerful tool that collects scattered coherent x-rays in the far-field on x-ray sensitive area detectors and uses phase retrieval algorithms to reconstruct the 3D electron density and lattice displacement fields in nanocrystals[13–17]. The penetrating power of high energy x-rays makes BCDI an ideal probe for studying in situ processes while the full 3D detail it provides aids in understanding the complex role of crystallographic facets, extended defects, and curvature in catalytic activity at the nanoscale[20,21].

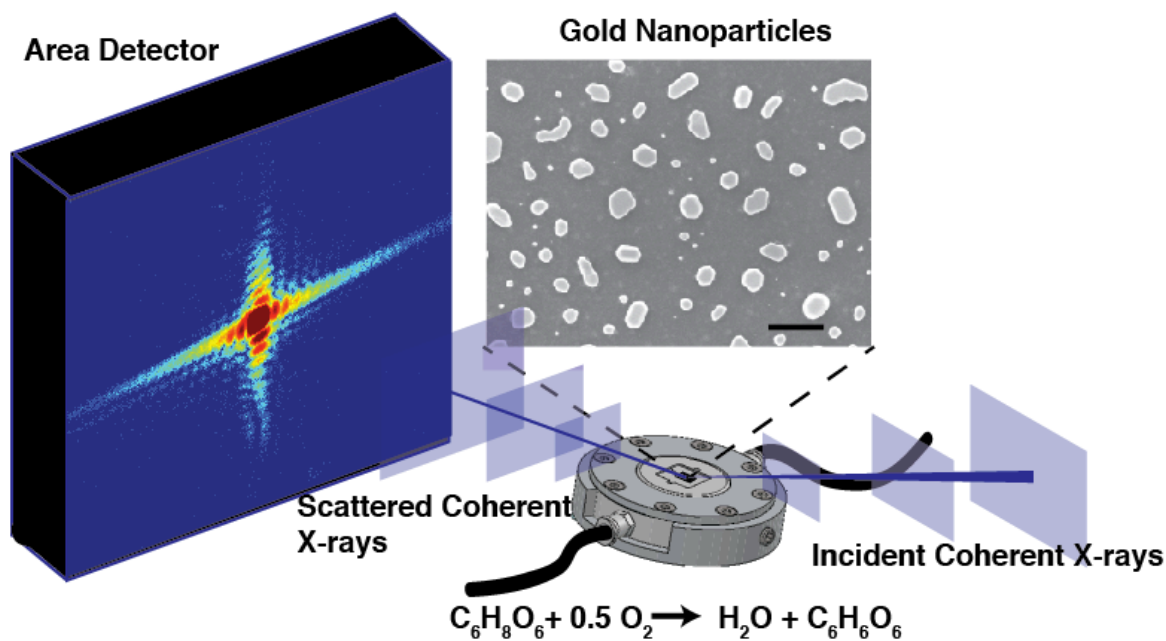

**Figure 1**. Imaging a single gold nanoparticle catalyst during ascorbic acid decomposition with coherent x-rays. Coherent x-rays are incident on a liquid cell containing gold nanoparticles on a silicon substrate (scale bar 1 micron). Different solutions are injected into the cell. Diffracted x-rays from a single gold nanoparticle are continuously collected on an x-ray sensitive area detector.

The experimental setup is shown schematically in **Figure 1**. Focused coherent x-rays are incident on a liquid environmental x-ray cell (see Figure S1) that contains gold nanoparticles on a silicon substrate (see Figure S2). A Mylar film controls the fluid layer thickness to smaller than 200 microns, as estimated from the reduction of scattered x-ray intensity when water was added to the cell. The gold nanoparticles have various shapes with sizes between 150-700 nm. The x-rays scattered by a single gold nanoparticle, satisfying the $(11\bar{1})$ Bragg condition, are recorded on an area detector in the different sample environments (see Methods)[22]. From the coherent diffraction data, we reconstruct both the 3D distribution of Bragg electron density[23], $\rho(x,y,z)$, and the 3D lattice displacement field projected along $[11\bar{1}]$, $u_{11\bar{1}}(x,y,z)$, with 17 nm spatial resolution as defined by the phase retrieval transfer function (see Figure S3) and 10 minute temporal resolution. Although the spatial resolution is 17 nm, the strain resolution is controlled by the strong sensitivity of x-rays to the crystal lattice spacing and is of the order of ~$10^{-4}$ [24].

**Figure 2** shows a particle that exhibits lattice dynamics during acid exposure. The particle is approximately 200x320x200 nm in size with a shape shown by the green semitransparent isosurface in Fig. 2a. While smaller nanoparticles are the most relevant to industrial catalysis, this nanoparticle size still scatters significantly while immersed in the solution. The red and blue isosurfaces in Fig. 2a are drawn at $u_{11\bar{1}}$ displacement values of +0.3 Å and -0.7 Å, respectively. As synthesized, the particle has two regions of -0.7 Å displacement in two corners. The 0.1M acid solution is injected into the cell at t = $0_+$. 10 minutes later, the negative displacement at the top right corner is reduced below the isosurface value leading to a disappearance of the region while the size of the other negative region is also decreased. Note that this does not mean the displacement in this region becomes zero. When the displacement field value at a given location is reduced below the isosurface value no isosurface is drawn. The particle also shows small regions of expansion at particular edges. At Δt=20m, the displacement at the lower left corner is relieved further while the displacement at the upper right corner begins to return to the isosurface value. The particle shows expansion at the edge where two flat facets meet as well as in a particular corner close to the substrate. At Δt=30m, only one expansive displacement region remains and the compressive regions return to the corners. Finally, at Δt=40m, the compressive displacement is in top right and bottom left corners, the positive displacement is below the isosurface value, and the particle's $u_{11\bar{1}}$ 3D displacement field is well correlated to its initial state (for a quantitative correlation analysis, see Figure S4).

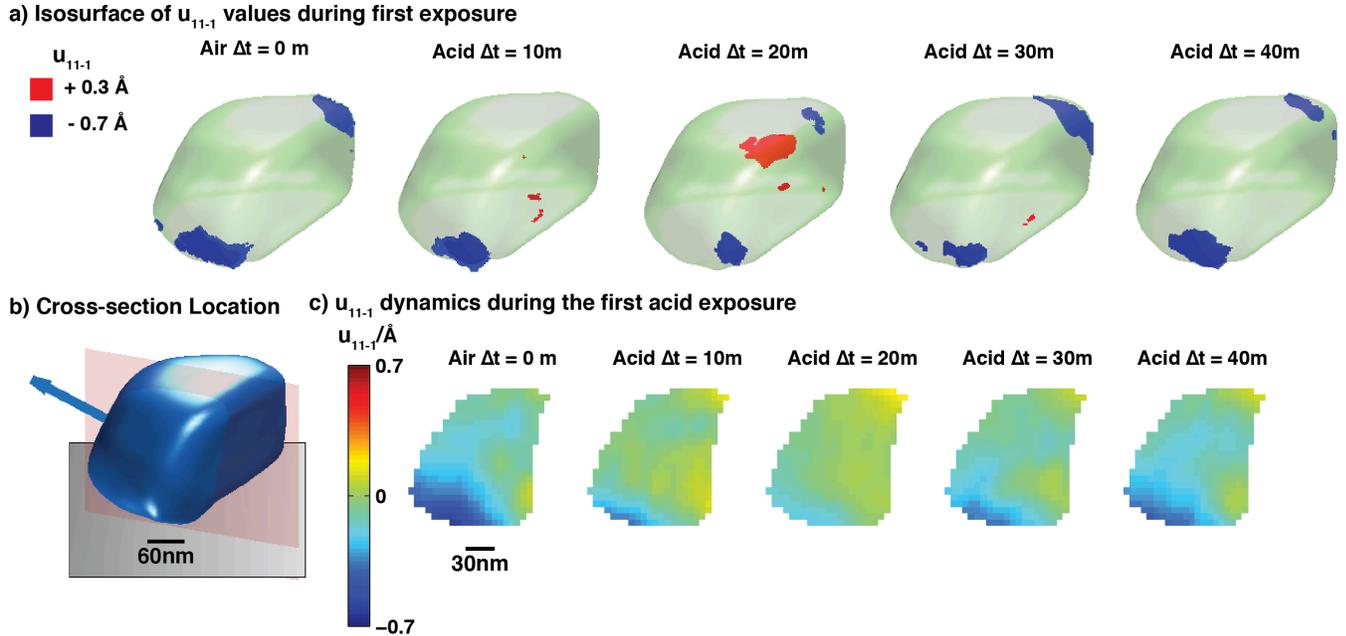

**Figure 2**. Displacement field dynamics during the first ascorbic acid exposure. **a**, The particle shape is shown as a semitransparent green isosurface. Blue and red regions indicate displacements along [11-1], $u_{11\bar{1}}$, of +0.3 Å and -0.7 Å, respectively, at particular times during ascorbic acid decomposition. **b**, The particle's shape is shown as a solid blue isosurface. The blue vector is the $Q_{11-1}$ scattering vector and the red slice plane is oriented such that **Q** is contained within the plane. **c**, The displacement field dynamics during the first ascorbic acid exposure at the cross-section indicated by the red plane in **b**.

Figure 2b shows the particle shape as a blue isosurface and the scattering vector, $\mathbf{Q}_{11-1}$, as a blue arrow. The red cross-section is oriented such that **Q** is contained within the plane. Figure 2c shows the displacement field dynamics at this cross-section as a function of time. One noticeable acid effect is the expansion in the bottom left and upper right crystal regions, similar to what is shown by the isosurface renderings in Fig. 2a. The particle was washed with water, dried, and exposed to a second 0.1M acid solution. See Figure S5 for the displacement field dynamics during the second acid exposure, which are similar to the first exposure. We also investigated two other crystals and summarize the results in Figure S6.

BCDI measurements reveal local lattice displacements of up to 0.3-0.4 Å during ascorbic acid exposure. These large displacements are primarily concentrated near the under-coordinated nanocrystal facets and edges. The under-coordinated sites of a nanocrystal are also expected to serve as preferential sites for adsorption of molecular species during a catalytic process. This indicates that images obtained *via* BCDI measurements can be used to identify not only what particles are catalytically active but also the active sites. This ability to systematically investigate single nanoparticle activity presents an opportunity to identify the optimal morphology and size for catalysts.

To investigate the origin of the strain at the atomic level and to uncover the reaction pathway, we perform RMD simulations using the ReaxFF formalism[25–27]. ReaxFF is proven for a diverse set of systems, including Au-S-C-H[28], hydride formation in palladium nanoclusters[29], and gold oxides[30]. For comparison between ReaxFF, embedded atom method, and density functional theory, see Table S1, Table S2, and Figure S7. We note that, due to computational limitations of RMD simulations, it is inevitable to have significant differences in scale between the simulation and experiment. Please see the Supplemental Material for further discussion of the limitations of RMD and the ReaxFF parameter set. The goal of the RMD simulations is to investigate if there are any atomic-level processes that can create large local displacements in the nanocrystal lattice during ascorbic acid decomposition and to provide atomistic insight for the reaction-related forces applied in the finite element analysis (FEA) model.

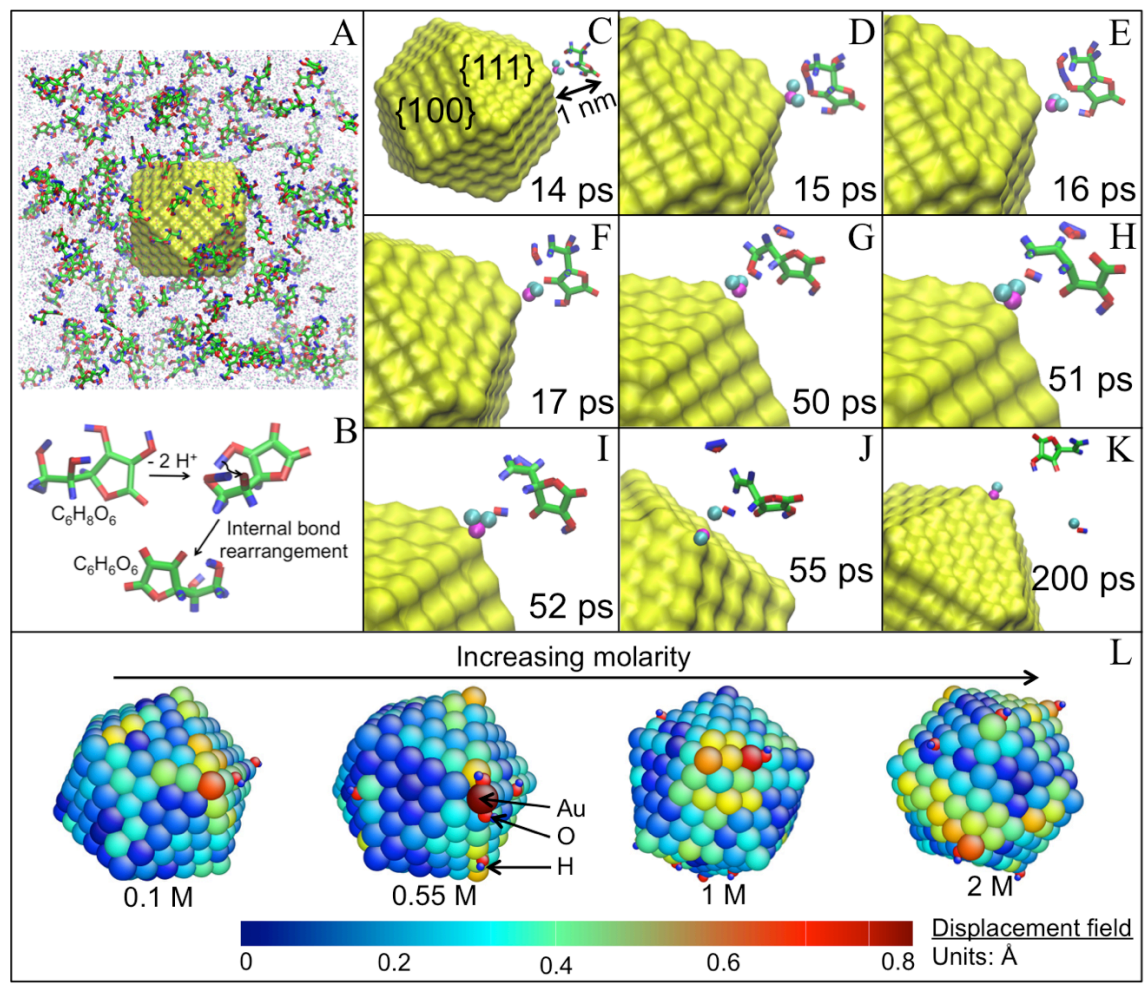

**Figure 3. Reactive molecular dynamics simulation results of gold-catalyzed ascorbic acid decomposition.** For clarity, only the trajectory of a representative ascorbic acid and water molecule are shown in c-k. **a**, Gold nanoparticle in a well-dispersed 1 M acid solution. **b**, Ascorbic acid decomposition pathway in the presence of dissolved oxygen. **c-k**, The OH$^-$ ion adsorption pathway at a corner site of a gold nanoparticle. The pathway involves co-physisorption of a water molecule and the acid, and dissociation of the adsorbed water molecule into H$^+$ and OH$^-$ ions aided by an oxygen atom on the acid molecule. The dissociated OH$^-$ ion chemically binds to the corner site that induces a strain on the gold lattice. **l**, t = 200 ps snapshots of the local displacement fields for different acid molarities. The gold atoms are colored based on the atomic displacement from an initial unstrained reference. Smaller blue and red spheres show
9

chemisorbed OH⁻ ions. Color scheme for a-k – yellow: gold, green: carbon, red: oxygen, blue: hydrogen, pink: oxygen in water, cyan: hydrogen in water.

**Figure 3** shows simulation results for a representative gold nanoparticle interacting with a 1 M ascorbic acid solution containing dissolved oxygen at 300 K. For further detail, see the supplementary material, Figure S8, and Table S3. Simulations were conducted at higher acid molarities than that of the experiments to enhance the kinetics of the reaction to time scales readily obtained on modest computational clusters (see Figure S9, Table S4-S6 for more information). The molarity primarily affects the kinetics and not the degree of atomic displacement induced. As a result of acid exposure, we observe active sites on both vertices and edges of the nanocrystal. In addition, we observe gold lattice strain only in the presence of both water and ascorbic acid (see supplementary material).

The RMD simulations indicate a sequence of co-adsorption and acid-aided dissociation of water on a gold nanocrystal, which in turn leads to local straining of the gold lattice. The process is initiated when water and ascorbic acid molecules co-physisorb at an under-coordinated site on the gold lattice (typically corners and edges of the nanocrystal facets, see Fig. 3C). The decomposed ascorbic acid (Fig. 3B, 3D-3F) facilitates the dissociation of the adsorbed water molecule (Fig. 3G-3J). The dissociated OH⁻ ion chemically binds to the gold nanoparticle (Fig. 3K) and induces local straining. Fig. 3L depicts a snapshot of the localized displacement field in the gold nanocrystal with reference to an unstrained gold nanocrystal lattice. At the active gold corner sites, where the OH⁻ ion chemisorbs, significant displacement fields are observed.

The RMD simulations reveal that the under-coordinated corners and edges of the nanocrystal are the preferred adsorption sites for molecular species. The results also indicate that formation of a strong chemical bond between Au and hydroxyl ions can cause significant local displacement fields. RMD demonstrates the connection between local strain and local activity. Although there are differences in scale due to technical limitations between the RMD simulation and the BCDI experiment, the two are complementary. Using FEA, we further investigate whether multiply hydroxyl ion chemisorptions could result in a displacement field similar to the one observed in the BCDI experiment.

**Figure 4** shows the results of the finite element simulation performed on the experimental nanocrystal reconstructed from the BCDI measurements. The force exerted by a hydroxyl ion on a gold atom during the formation of a chemical bond was determined from the RMD simulation to be $1.76 \times 10^{-10}$ N. Assuming a 2% coverage of hydroxyl ions (see supplementary material for more details) on a 612.9 nm$^2$ area corner patch (having 13.87 gold atoms per nm$^2$), a total force $F_{MD}$ is obtained as $3 \times 10^{-8}$ N. This force is applied to the patch of the nanocrystal while the mesh points near the opposite corner are subject to the Dirichlet constraint of zero displacement. The steady state FEA solution for this system revealed a maximum $u_{11\text{-}1}$ projected displacement of 0.35 Å (consistent with the experimental observation in Fig. 2C). In addition, the steady state displacement field extends 40-60 nm from the corner into the particle interior, which is comparable to the experimental observations in Fig. 2 and Fig. S5.

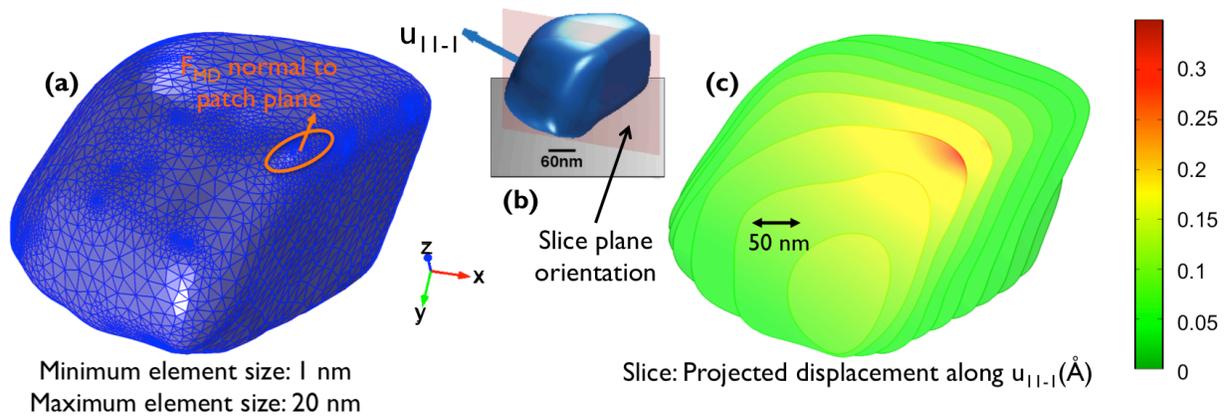

**Figure 4.** Finite element analysis on the experimental nanocrystal. **a**, Force derived from molecular dynamics ($F_{MD}$) is applied on a corner patch of a gold nanocrystal while the mesh points near the opposite corner are subject to the Dirichlet constraint of zero displacement. **b**, Schematic showing the crystallographic [1 1 -1] direction with respect to the Cartesian frame and the orientation of the slice planes shown in **c**. **c**, The resulting displacement field projected onto the [11-1] direction shown on several slice planes.

In conclusion, we studied 3D lattice displacement dynamics in gold nanoparticles during ascorbic acid decomposition experimentally using Bragg coherent diffractive imaging (BCDI) and theoretically using reactive molecular dynamics (RMD) and RMD-aided finite element analysis (FEA). The displacement field changes observed via BCDI during acid exposure are related to chemical interactions via RMD simulations. RMD simulations offer a novel mechanism to understand the observed lattice dynamics: acid-aided dissociation of water and subsequent chemisorption that causes local atomic displacements. The FEA results indicate that the results from the RMD simulations performed on small nanocrystals are transferable to larger crystal sizes. Our results show the utility of imaging the 3D displacement field evolution to identify catalytically active nanoparticles due to the fact that they exhibit strain during redox catalysis reactions. This identification can be carried out at the single particle level under operating conditions and serves as an unambiguous method to identify the optimal morphology and size for catalysts.

ASSOCIATED CONTENT
**Supporting Information available**: Sample synthesis, phase retrieval algorithms, and reactive molecular dynamics are described in additional detail in the supporting information. Eleven (11) additional figures and six (6) tables are provided. This material is available free of charge *via* the Internet at http://pubs.acs.org.

AUTHOR INFORMATION

**Notes**

The authors declare no competing financial interest.

**Author Contributions**

A.U., R.H., and J.W.K. conducted the experiment. A.U. performed the experimental data analysis. J.N.C. provided samples. SKRS guided the simulation effort. K.S and S.A.D performed the reactive MD simulations. B.N performed the electronic structure simulations. K.S. and R.H. performed the finite element analysis. K.S., S.A.D., B.N., N.F. and SKRS performed the simulation data analysis. All authors interpreted the results and wrote/revised the manuscript. A.U. and P.M. conceived the experiment.

**Acknowledgements:**

This work was supported by U.S. Department of Energy, Office of Science, Office of Basic Energy Sciences, under Contract DE-SC0001805. OGS and AU are grateful to the UCSD Inamori Fellowship. JNC gratefully acknowledges financial support from the Volkswagen Foundation. PM thanks the ARC for support through LF100100117. This research used resources of the Center for Nanoscale Materials and the Advanced Photon Source, a U.S. Department of Energy (DOE) Office of Science User Facility operated for the DOE Office of Science by Argonne National Laboratory under Contract No. DE-AC02-06CH11357. We thank the staff at Argonne National Laboratory and the Advanced Photon Source for their support.

ABBREVIATIONS

RMD, reactive molecular dynamics; SEM, scanning electron microscopy; BCDI, bragg coherent diffractive imaging; FEA, finite element analysis.